\documentclass{fmj2010}
\usepackage{graphicx}
\usepackage{natbib}

\def\fermilat{\textit{Fermi}/LAT}

\begin{document}
   \title{The Jet in M87 from e-EVN Observations}

   \author{G. Giovannini\inst{1,2}
          \and
          C. Casadio\inst{1,2}
          \and
          M. Giroletti\inst{1}
          \and
          M. Beilicke\inst{3}
          \and
          A. Cesarini\inst{4}
          \and
          H.  Krawczynski\inst{3}
          }

   \institute{Istituto di Radioastronomia-INAF, via Gobetti 101, 40129 Bologna, Italy         \and
             Dipartimento di Astronomia, via Ranzani 1, 40127 Bologna, Italy
             \and
             Department of Physics, Washington University, St. Louis, MO 63130, USA
             \and
             School of Physics, National University of Ireland Galway, University Road, Galway, Republic of Ireland
             }

   \abstract{ One of the most intriguing open questions of today's astrophysics
     is the one concerning the location and the mechanisms for the production
     of MeV, GeV, and TeV gamma-rays in AGN jets. M87 is a privileged
     laboratory for a detailed study of the properties of jets, owing to its
     proximity, its massive black hole, and its conspicuous emission at radio
     wavelengths and above. We started on November 2009 a monitoring program
     with the e-EVN at 5 GHz, during which two episodes of activity at energy
     E$>$100 GeV have occured. We present here results of these multi-epoch
     observations. The inner jet and HST-1 are both detected and resolved in
     our datasets. One of these observations was obtained at the same day of
     the first high energy flare. A clear change in the proper motion velocity
     of HST-1 is present at the epoch $\sim$ 2005.5. In the time range 
     2003 -- 2005.5 the apparent velocity is subluminal, and superluminal
     ($\sim$ 2.7c) after 2005.5. }

   \maketitle
%

\section{Introduction}

The giant radio galaxy Messier 87 (M87), also known as 3C 274 or Virgo A, is
one of the best studied radio sources and a known 
$\gamma$-ray-emitting AGN.  It
is located at the center of the Virgo cluster of galaxies at a distance = 16.7
Mpc, corresponding to an angular conversion 1 mas = 0.081 pc. The massive black
hole at the M87 center has an estimated mass = 6 $\times$ 10$^9$ solar masses,
with a scale of 1 mas = 140 R$_S$. The bright jet is well resolved in the
X-ray, optical, and radio wave bands.

 \begin{figure}
   \centering
 \includegraphics[width=8.5cm]{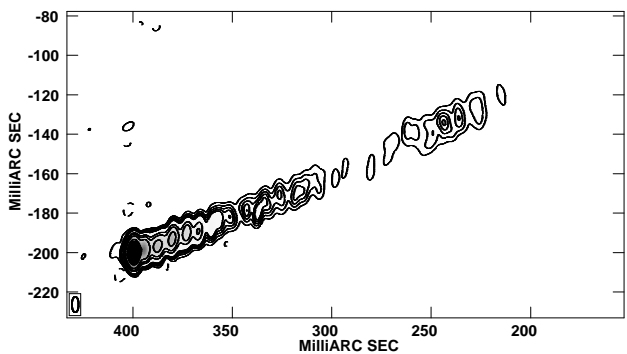}
      \caption{e-EVN image at 5 GHz of the inner jet of M87, 
February 10th, 2010
epoch. The HPBW is 
   8 $\times$ 3.4 mas. Levs are -2 2 4 6 8 10 30 50 100 300 500 1000 1500 
       mJy/beam}
         \label{fig1}
   \end{figure}

The jet is characterized by many substructures and knots.  In 1999 HST
observations revealed a bright knot at about 1'' from the core, named
HST-1. This feature is active in the radio, optical, and X-ray regimes.  It was
discussed by \citet{Perlman1999}, who compared optical and radio images.
\citet{Biretta1999} measured in the range 1994-1998 a subluminal speed = 0.84c
for the brightest structure (HST-1 East), which appears to emit superluminal
features moving at 6c. However this motion was measured in regions on a 
larger scale with respect to the VLBI structures discussed here.

VLBI observations of the M87 inner region show a well resolved, edge-brightened
jet structure. At very high resolution (43 and 86 GHz) near to the brightest
region the jet has a wide opening angle, and we refer to the many published
papers which discuss the possible presence of a counter-jet and the location of
the radio core \citep[see e.g.][]{Junor1999,Krichbaum2005,Ly2007,Kovalev2007}.

Very High Energy (VHE) $\gamma$-ray emission was reported by the High Energy
Gamma-Ray Astronomy (HEGRA) collaboration in 1998/99 \citep{Aharonian2003},
confirmed by the High Energy Stereoscopic System (HESS) in 2003-2006
\citep{Aharonian2006}, and by VERITAS in 2007 \citep{Acciari2008}. Coordinated
intensive campaigns have permitted to detect the source again in 2008
\citep{Acciari2009} and as recently as February and April 2010
\citep{Mariotti2010,Ong2010}. Steady emission at MeV/GeV energies has also 
been detected by \fermilat\ \citep{Abdo2009}.

Various models have been proposed to explain the multi-wavelength emission and
in particular to constrain the site of the VHE emission in M87. The
inner jet region was favoured by the observed short TeV variability timescales
\citep{Aharonian2006}. The VHE emission could then be produced in
the BH magnetosphere \citep{Neronov2007} or in the slower jet layer
\citep{Tavecchio2008}, with the spine accounting for the emission from
the radio to the GeV band; this would lead to a complex correlation between the
TeV component and the lower energy ones.

However, VLBA observations at 1.7 GHz \citep{Cheung2007} resolved HST-1 in
substructures with superluminal components.  \citet{Aharonian2006}
discussed that HST-1 cannot be excluded as a source of TeV $\gamma$ rays, 
however they conclude that the more promising possibility is that the site of 
TeV $\gamma$-ray production is the nucleus of M87 itself. Comparing 
multifrequency data \citet{Harris2008} suggested that the TeV emission 
from M87 was originated in HST-1.

Finally, \citet{Acciari2009} reported rapid TeV flares from M87 in February
2008, associated by an in increase of the radio flux from the nucleus, while
HST-1 was in a low state, thus concluding that the TeV flares originate in the
core region.

In this context we started at the end of 2009 a program to observe with the 
e-EVN
M87 at 5 GHz to study the properties of the M87 core, jet, and HST-1 structure.

\section{Observations and Data Reduction}

Our original monitoring schedule included four epochs at 5 GHz, to be carried
out before and during the season of visibility from the TeV telescopes, namely
on 2009 November 19, 2010 January 27, February 10, and March 28.  Following the
February \citep{Mariotti2010} and April \citep{Ong2010} high energy events,
three more epochs have been added on 2010 March 6, May 25, and June 9 as Target
of Opportunity (ToO) observations.

The observations have been carried out in eVLBI mode, with data acquired by 
EVN
radio telescopes, directly streamed to the central data processor at JIVE, and
correlated in real-time. The observing frequency of 5 GHz was chosen to
simultaneously grant a large field of view and a high angular resolution. For
observations taking advantage of the long baselines provided by the Arecibo and
Shanghai telescopes, our clean beam with uniform weights is about $2.0 \times
0.9$ mas in PA $-25^\circ$.

As a result of the large bandwidth (a rate of 1 Gbps was sustained by most 
stations),
long exposure (up to 6 hours per epoch), and extended collecting area, the rms
noise in our images is mostly dynamic range limited. As an average value, we
can quote 0.5 -- 0.8 mJy~beam$^{-1}$ in the nuclear region and 0.1 -- 0.2 
mJy~beam$^{-1}$ in the HST-1 region. 
We present here preliminary results from the first
5 epochs. The reduction of May 25th data is still in progress.

\section{Results}

\subsection{The inner jet region}

In Fig.~\ref{fig1} we show the image obtained on Feb. 10th, 2010 of the 
core and inner jet
region of M87. The jet orientation and velocity has been discussed
in many papers comparing observational data on the jet brightness and
proper motion. Recently \citet{Acciari2009} assumed as a likely range 
$\theta$ = 15 -- 25 deg.

\begin{figure}
   \centering
 \includegraphics[width=8.5cm]{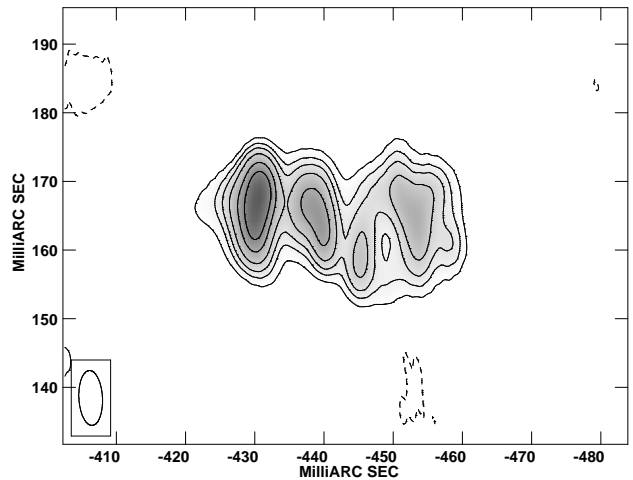}
      \caption{e-EVN image at 5 GHz of HST-1, January 27th, 2010 epoch. 
The HPBW is
   8 $\times$ 3.4 mas. Levs are -0.3 0.3 0.5 0.7 1 1.5 2 3
       mJy/beam}
         \label{fig2}
   \end{figure}

\begin{figure}
   \centering
\includegraphics[width=8cm]{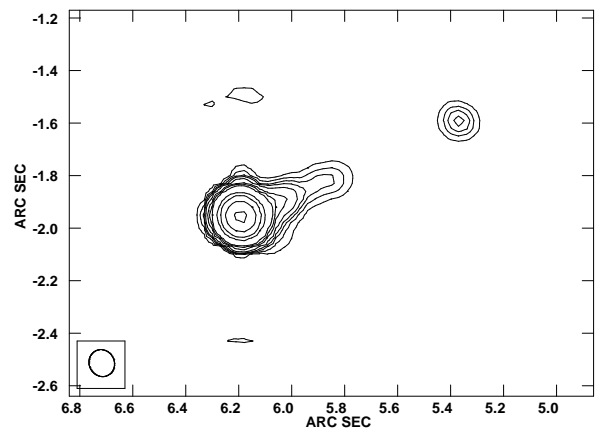}
      \caption{VLA image at 15 GHz of the inner jet of M87 obtained 
on November 2008. The HPBW is 0.10 arcsec.
   Levs are 5 10 15 20 30 50 70 100 300 500 1000 2000
       mJy/beam}
         \label{fig3}
   \end{figure}

\begin{table}
\begin{minipage}[t]{\columnwidth}
      \caption[]{e-EVN Results}
         \label{table1}
\centering
         \begin{tabular}{c c c c}
            \hline\hline
            Epoch & S(core) & HST-1 peak & Distance \\
              yr  &  mJy       &   mJy/beam      & mas      \\
            \hline
            2009.87 & 1806 & 3.5 & 905.6 \\
            2010.07 & 1810 & 2.7 & 907.3 \\
            2010.11 & 1798 & 3.0 & 909.5 \\
            2010.17 & 1891 & 4.6 & 911.6 \\
            2010.25 & 2013 & 3.4 & 912.4 \\
\noalign{\smallskip}
            \hline
\multicolumn{3}{l}{\scriptsize S$_{core}$ uncertainty is $\sim$ 30 mJy;}\\
\multicolumn{3}{l}{\scriptsize Distance is between the HST-1 peak flux and 
the core,}\\
\multicolumn{3}{l}{\scriptsize  with an uncertainty of $\sim$ 2 mas}\\
         \end{tabular}
         \end{minipage}
         \end{table}

We used our images to search for evidence of a 
possible proper motion, comparing different epoch position of jet
substructures and subtracting images at different epochs (with the
same grid, angular resolution and similar uv-coverage) to look for
possible systematic trends. No evidence was found in anycase.

\begin{figure*}
   \centering
\includegraphics[width=16.5cm]{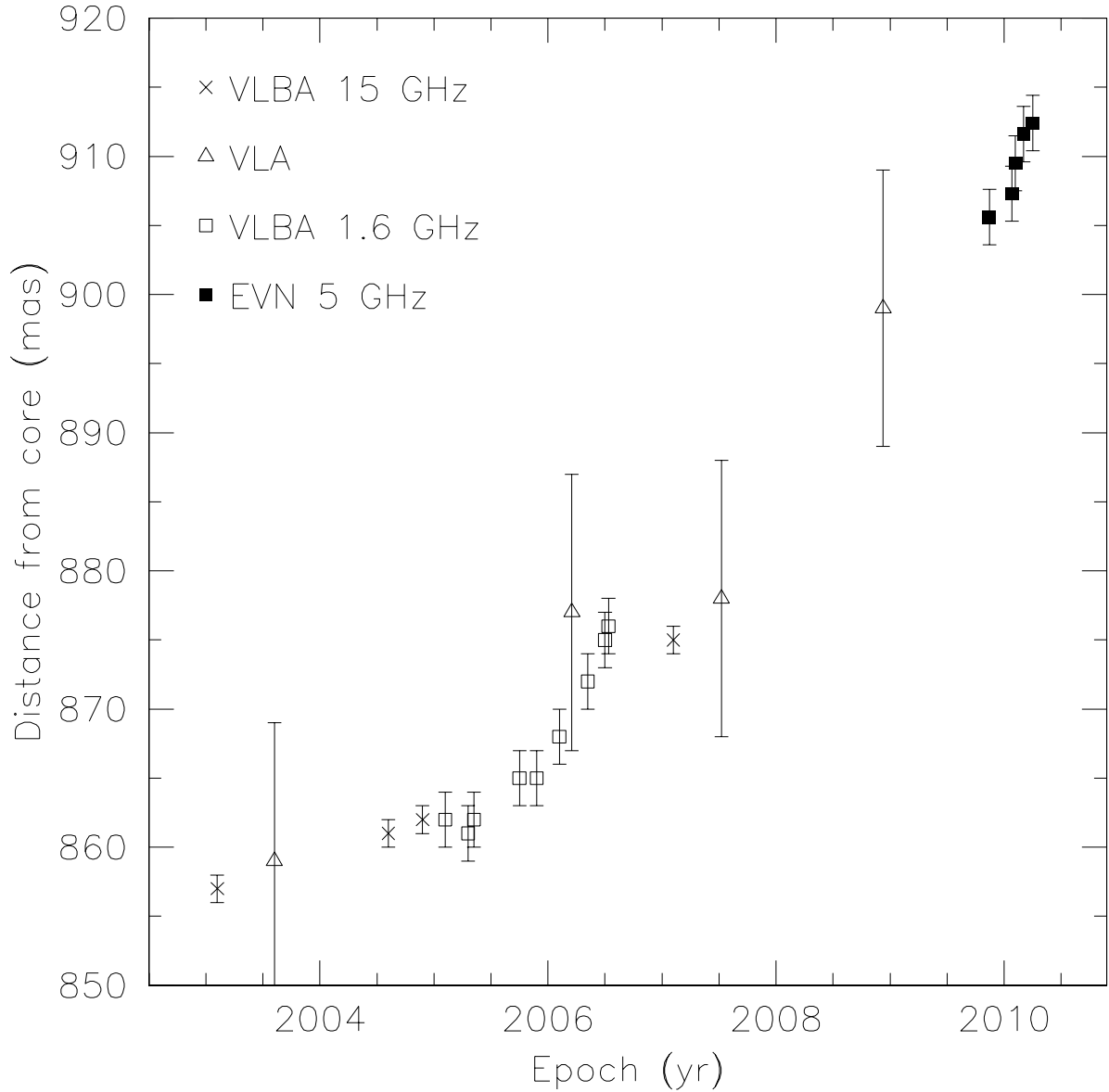}
      \caption{Distance of HST-1 brightest peak from the M87 core at
      different epochs}
         \label{fig4}
   \end{figure*}

From the jet/counter-jet brightness ratio we derive that 
$\beta cos\theta$ $>$ 0.82 which implies $\theta$ $<$ 35$^\circ$.

To better constrain the jet velocity and orientation we assume that the jet 
limb-brightened structure is due to a velocity structure: a fast inner spine
and a relatively slow external shear layer. In a range of $\theta$ it is 
possible
that the Doppler factor of the slower external regions is higher than the
Doppler factor of the fast spine. Therefore the apparent brightness of the 
spine can appear
fainter than that of the shear layer. We estimated different
Doppler factors using reasonable velocities and jet orientation. We found
that if the orientation angle $\theta$ is in the range 
20$^\circ$ -- 30$^\circ$ the
limb-brightened structure can be due to a different Doppler factor.

In  Table~\ref{table1}
we report the nuclear flux density measured in images at the same angular 
resolution (8 $\times$ 3.4 mas). We find a 
marginal evidence of a nuclear flux density increasing in the last two epochs,
and of a high HST-1 peak flux on 2010.17. 

\subsection{HST-1}

In our observations HST-1 is clearly resolved (Fig.~\ref{fig2}). 
Comparing different epochs
we are able to individuate the brightest knot and to measure its proper
motion (Table~\ref{table1}). 
Other features appear to move at about the same velocity, 
but because
of the low brightness and complex morphology we discuss here only the position
of the brightest knot inside HST-1.

To better study the dynamic of this structure we searched archive VLA data at 
high resolution (A configuration) and high frequency (X, U, and Q bands).
We refer to \citet{Harris2009} for a discussion of the flux density 
variability. Here we only want to compare different epochs to derive the HST-1 
dynamic.

We started to analyze data from 2003.6 since, as shown in \citet{Harris2009},
in previous epochs few data are availble, moreover the HST-1 flux density was 
very low. A better analysis with a larger time range will be presented in
a future paper.

Starting from 2003.6 the HST-1 structure
is well evident (see e.g. Fig.~\ref{fig3} obtained on November 2008) and 
well separated by the jet structure near the 
core. 
From VLA multifrequency observations at the same time we
find that the total spectrum for HST-1 is moderately steep: on 2006.21 is 
$\alpha^{8.4}_{15}$ = 0.68 and $\alpha^{15}_{22}$ = 0.82
The high frequency steepening is not always present, 
on 2007.52 $\alpha^{8.4}_{15}$ is the same but  $\alpha^{15}_{22}$ = 0.36.
This variability and trend of the radio spectrum is in agreement with the 
substructures and 
variability visible in VLBI images of HST-1. 

We estimated from e-EVN and VLA data the distance of HST-1 from the core.
In e-EVN data we measured the distance between the core and the brightest
knot in HST-1, in VLA images we used the HST-1 peak, being 
this structure unresolved. Adding the values obtained at 1.5 and 15 GHz
by \citet{Cheung2007} and \citet{Chang2010}, respectively, 
we can study the HST-1 proper motion with a good statistic from 2003 to 
present epoch. The apparent proper motion of HST-1 is shown in  
Fig.~\ref{fig4}.
Unfortunatelly in the time range 2006.5 -- 2007.5 there are only a few points
not completely in agreement, however taking into account the different 
observing frequency, angular resolution
and radio telescopes, the proper motion looks well defined.

A clear change in the proper motion velocity is present at the epoch 
$\sim$ 2005.5, coincident with the TeV $\gamma$-ray activity and
the maximum radio/X-ray flux density of
HST-1. In the time range 2003 -- 2005.5 the apparent velocity is 
0.5c -- 0.6c; in the time range 2005.5 -- 2010.25 the apparent velocity is
$\sim$ 2.7c.
Assuming a jet orientation angle = 25$^\circ$ a proper motion of 2.7c 
corresponds to an intrinsic velocity = 0.94c.

\section{Summary}

With our new e-EVN data obtained in the time range 2009.87 -- 2010.25 
we have obtained images of the nuclear region of M87 and of the jet 
substructure HST-1.

The radio core flux density is constant in the first three epochs 
with an average flux density $\sim$ 1805 mJy and slightly increasing in the 
last two epochs: 2013 mJy in 2010.25.

The inner jet is transversally resolved and assuming that the limb-brightened
structure is due to a different Doppler factor because of the jet velocity
structure we derive a source orientation angle $\theta$ = 20 -- 30 degree. 

The HST-1 structure is well resolved in many substructures. A proper motion
is clearly present. Comparing e-EVN data with archive VLA data and published
VLBA data at 1.7 and 15 GHz we find a strong evidence that in 2005.5
HST-1 increased its velocity from an apparent velocity = 0.5 -- 0.6 c 
to 2.7c. With present data it is not possible to discuss if this change in
velocity is related to the M87 VHE activity and/or to the maximum radio/X-ray 
flux density of HST-1 at this
epoch. A more regular and longer 
monitor and a multi-frequency comparison is necessary to clarify this point.

\begin{acknowledgements}
The European VLBI Network is a joint facility of European, South
African, and Chinese 
radio astronomy institutes funded by their national research councils.
The
National Radio Astronomy Observatory is operated by Associated Universities,
Inc., under cooperative agreement with the National Science Foundation.

\end{acknowledgements}


\begin{thebibliography}{}

\bibitem[Abdo et al.(2009)]{Abdo2009} Abdo, A.~A., Ackermann, M., Ajello, 
M., et al.\ 2009, ApJ, 707, 55 

\bibitem[Acciari et al.(2009)]{Acciari2009} Acciari, V.~A., Aliu, E., 
Arlen, T., et al.\ 2009, Sci, 325, 444 

\bibitem[Acciari et al.(2008)]{Acciari2008} Acciari, V.~A., Beilicke, M., 
Blaylock, G., et al.\ 2008, ApJ, 679, 397 

\bibitem[Aharonian et al.(2003)]{Aharonian2003} Aharonian, F., 
Akhperjanian, A., Beilicke, M., et al.\ 2003, A\&A, 403, L1 

\bibitem[Aharonian et al.(2006)]{Aharonian2006} Aharonian, F., 
Akhperjanian, A.~G., Bazer-Bachi, A.~R., et al.\ 2006, Sci, 314, 1424 

\bibitem[Biretta et al.(1999)]{Biretta1999} Biretta, J.~A., Sparks, W.~B., 
\& Macchetto, F.\ 1999, ApJ, 520, 621 

\bibitem[Chang et al.(2010)]{Chang2010} Chang, C.~S., Ros, E., Kovalev, Y.~Y., Lister, M.~L. 2010, A\&A in press (arXiv:1002.2588)

\bibitem[Cheung et al.(2007)]{Cheung2007} Cheung, C.~C., Harris, D.~E., 
Stawarz, L., 2007, ApJ, 663, L65

\bibitem[Harris et al.(2008)]{Harris2008} Harris, D.~E., Cheung, C.~C., 
Stawarz, L., \& al.\ 2008, in ASP Conf. Ser. 386, 80

\bibitem[Harris et al.(2009)]{Harris2009} Harris, D.~E., Cheung, C.~C.,
Stawarz, L., Biretta, J.~A., Perlman, E.~S. 2009, ApJ, 699, 305

\bibitem[Junor et al.(1999)]{Junor1999} Junor, W., Biretta, J.~A., 
\& Livio, M.\ 1999, Natur, 401, 891 

\bibitem[Kovalev et al.(2007)]{Kovalev2007} Kovalev, Y.~Y., Lister, M.~L., 
Homan, D.~C., \& Kellermann, K.~I.\ 2007, ApJ, 668, L27 

\bibitem[Krichbaum et al.(2005)]{Krichbaum2005} Krichbaum, T.~P., Zensus, 
J.~A., \& Witzel, A.\ 2005, AN, 326, 548 

\bibitem[Ly et al.(2007)]{Ly2007} Ly, C., Walker, R.~C., Junor, W.\ 2007,
ApJ, 660, 200

\bibitem[Marconi et al.(1997)]{Marconi1997} Marconi, A., Axon, D.~J., 
Macchetto, F.~D., Capetti, A., Sparks, W.~B., 
\& Crane, P.\ 1997, MNRAS, 289, L21 

\bibitem[Mariotti(2010)]{Mariotti2010} Mariotti, M.\ 2010, ATel, 2431, 1 

\bibitem[Neronov \& Aharonian(2007)]{Neronov2007} Neronov, A. \& Aharonian, 
F.~A., 2007 ApJ 671, 85

\bibitem[Ong(2010)]{Ong2010} Ong, R.~A.\ 2010, ATel, 2443, 1 

\bibitem[Perlman et al.(1999)]{Perlman1999} Perlman, E.~S., Biretta, J.~A., 
Zhou, F., Sparks, W.~B., \& Macchetto, F.~D.\ 1999, AJ, 117, 2185 

\bibitem[Tavecchio \& Ghisellini(2008)]{Tavecchio2008} Tavecchio, F. \& 
Ghisellini, G. 2008, MNRAS, 385, 98

\end{thebibliography}
\end{document}